\documentclass[twocolumn,showpacs,preprintnumbers,amsmath,amssymb]{revtex4-1}

\usepackage{graphicx}
\usepackage{amssymb}
\usepackage{hyperref}

\begin{document}

\title{Exploring high temperature magnetic order in CeTi$_{1-x}$Sc$_x$Ge}

\author{J.G. Sereni, P. Pedrazzini, M. G\'omez Berisso, A. Chacoma, S. Encina}

\address{Low Temperature Division, CAB-CNEA and CONICET, 8400 San Carlos de Bariloche, Argentina}

\author{T. Gruner, N. Caroca-Canales, C. Geibel}

\address{Max-Planck Institute for Chemical Physics of Solids, D-01187
Dresden, Germany}

\date{\today}

\begin{abstract}

{Most of magnetic transitions related to Ce ordering are found
below $T_{ord} \approx 12$K. Among the few cases exceeding that
temperature, two types of behaviors can be distinguished. One of
them is related to the rare cases of Ce binary compounds formed in
BCC structures, with a quartet ground state, whose degeneracy (N =
4) is reduced by undergoing different types of transitions mostly
connected with structural modifications. The other group shows
evidences of itinerant character with the outstanding example of
CeRh$_3$B$_2$ showing the highest ordering temperatures $T_{ord} =
115$K. The second highest ordering temperature has been reported
for CeScGe with $T_{ord} = 47$K, but the nature of this magnetic
state has not been investigated very deeply. In order to shed more
light into this unusual high temperature ordering we studied the
structural, magnetic, transport and thermal properties of
CeTi$_{1-x}$Sc$_x$Ge alloys in the stability range of the
CeScSi-type structure $0.25 \leq x \leq 1$ This system presents a
rich variety of magnetic behaviors along this concentration range,
with the magnetic ordering growing from ferromagnetic (FM) $T_C
\approx 7$K up to an antiferromagnetic (AFM) transition at $T_N =
47$K. The different regions show the following characteristics:

i) on the Ti rich side ($0.25 \leq x \leq 0.50$) it exhibits a FM
ground state (GS) with large saturation magnetization values
$M_{sat}$ up to $\approx 1.15 \mu_B$. ii) Around $x = 0.60$, the
first crystal electric field excited doublet starts to contribute
to the GS magnetic properties. Furthermore an AFM component with a
connected metamagnetic transition appears. iii) At $x = 0.65$ a
clear change in the GS nature is associated to a critical point
above which the GS properties can be described like for an
itinerant system (with decreasing $M_{sat}$) and an effective GS
degeneracy $N_{eff} = 4$. iv) For $x > 0.65$, the magnetic phase
boundary splits into two transitions, with an intermediate phase
presenting incommensurate spin density waves features.}

\end{abstract}


\maketitle

\section{Introduction}

Magnetic transitions related to Ce ordering can be found along
four decades of  temperature, from 1.6mK in CMN \cite{CMN} and
115K in CeRh$_3$B$_2$ \cite{Malik}, though most of transitions
occur between ~1K and ~50K \cite{Bauer91,Hand91}. Different
magnetic behaviors characterize different temperature ranges of
magnetic order. In a simple description three groups can be
recognized: i) the one dominated by quantum fluctuations at $T\leq
2$K \cite{PhilMag}, ii) the classical behavior related to
localized Ce-4f moments mostly observed between $3 < T < 12$K
\cite{Bauer91}, and iii) the few cases whose ordering temperatures
exceed $T_{ord} \approx 12$K. Among these compounds, two types of
behaviors can also be distinguished. One of them is related to the
rare cases of Ce binary compounds formed in BCC structures, with a
quartet ground state (GS), whose degeneracy (N = 4) is reduced by
undergoing different types of transitions mostly connected with
structural modifications \cite{Hand91}. The other group shows
evidences of itinerant character \cite{DeLong91} with the
outstanding example of CeRh$_2$Si$_2$ showing one of the highest
ordering temperatures $T_{ord}=36$K \cite{Godart83}. Within the
large number of studies performed on this compound
\cite{CeRh2Si2}, evidences for both local and itinerant magnetic
character are equally claimed by different authors
\cite{8kawa,5settai97}. Such ambiguity is discussed since a long
time and recognized as the local-itinerant dilemma of Ce-$4f^1$
electrons \cite{Mackin}, not completely elucidated yet. This is
partially due to the lack of exemplary systems clearly running
between those limits driven by external control parameters like
alloying or pressure.

With the aim to contribute with further experimental information
to this unsolved problem, we have studied CeTi$_{1-x}$Sc$_x$Ge
alloys whose magnetic transition temperatures cover an unusually
extended range of temperature from $T_{ord} \approx 7$K in
CeTi$_{0.75}$Sc$_{0.25}$Ge, up to the second highest ordering
temperature among Ce compounds at $\approx 47$K of the
stoichiometric limit CeScGe \cite{Canfield91}. Within a large
range of thermal energy, the excited crystal-electric-field (CEF)
levels progressively contribute to the formation of the ordered
phase with the consequent change of the GS magnetic properties. A
crystalline modification from CeScSi- to CeFeSi-type structure
limits the range of study to $x \leq 0.25$.

\section{Experimental details and results}

Polycrystalline samples of CeTi$_{1-x}$Sc$_x$Ge with $0.25 \leq x
\leq 1$ were synthesized by arc melting under argon atmosphere the
nominal amounts of the constituents (purity above 99.99\%)
weighted inside an Ar atmosphere glove-box. The samples were
turned over and remelted several times to ensure homogeneity.
Then, the samples were placed in a tungsten boat wrapped with
zirconium foil and annealed at 1200\,$^o$C for one week. The
quality of the samples was verified by means of X-ray
powder-diffraction measurements using Cu-K$\alpha_1$ radiation
($\lambda =1.54056\, \rm{\AA}$) in a Stoe-Stadip-MP
diffractometer. The pattern was indexed on the basis of the
tetragonal CeScSi-type structure. Eleven samples were studied all
along the concentration range.

Specific heat was measured between 0.5 and 50K using a standard
heat pulse technique in a semi-adiabatic He$^3$ calorimeter, at
zero and applied magnetic field of $H = 40$kOe in some selected
samples. The magnetic contribution $C_m$ is obtained by
subtracting the phonon contribution extracted from
LaTi$_{0.5}$Sc$_{0.5}$Ge. DC-magnetization measurements were
carried out using a Quantum Design MPMS magnetometer operating
between 2 and 300K, and as a function of field up to 50kOe.
Electrical resistivity was measured between 2\,K and room
temperature using a standard four probe technique with an LR700
resistive bridge.

\subsection{Crystal structure}

\begin{figure}
\begin{center}
\includegraphics[angle=0, width=0.5 \textwidth, height=1.2\linewidth] {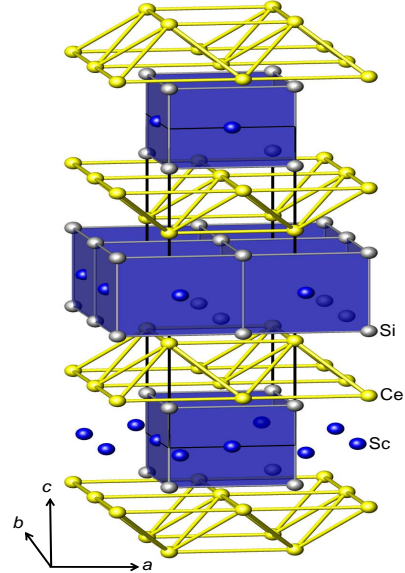}
\end{center}
\caption{(Color on line) CeScSi-type strucutre, showing double Ce
layers (yellow open network) and ligands layers (blue full
network).} \label{F1}
\end{figure}

The CeTi$_{1-x}$Sc$_x$Ge system forms in two related crystal
structure: CeFeSi-type for low Sc content (up to $x=0.15$) and
CeScSi-type beyond $x=0.23$  \cite{Gruner}. In the latter, each
second Ce-double layer is shifted by (1/2, 1/2) with a
rearrangement of Si atoms, becoming a body centered tetragonal
instead of primitive tetragonal of CeFeSi, see Fig.~\ref{F1}, with
the consequent doubling of the 'c' lattice parameter though the
first coordination spheres of Ce are identical.

\begin{figure}
\begin{center}
\includegraphics[angle=0, width=0.5 \textwidth, height=1.4\linewidth] {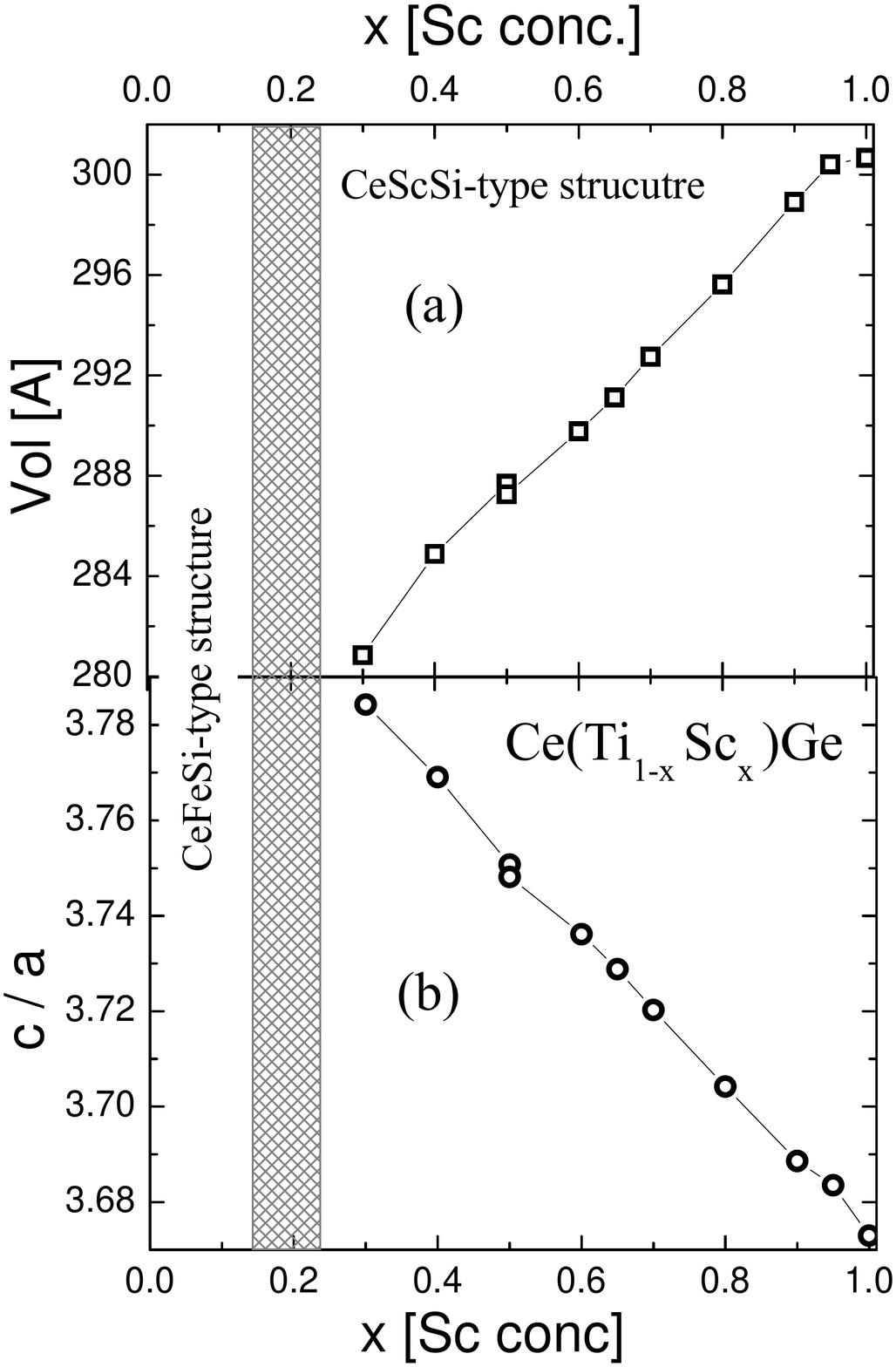}
\end{center}
\caption{(Color on line) (a) Unit cell volume variation in
CeTi$_{1-x}$Sc$_x$Ge as a function of Sc content. (b) c/a lattice
parameters ratio. The shaded area indicates the coexistence
region.} \label{F2}
\end{figure}

The unit cell volume dependence on Sc concentration and the 'c/a'
lattice parameters ratio are shown in Fig.~\ref{F2}a and b
respectively. The increase of the unit cell volume can be
explained by the larger atomic volume of Sc with respect to Ti,
whereas the reduction of the 'c/a' ratio is due to the increase of
the 'a' parameter because the 'c' one remains practically
unchanged. This indicates that an expansion in the basal plain of
the tetragonal structure occurs whereas the atomic distances in
the 'c' axis direction are not affected.

\subsection{Magnetic susceptibility}

\begin{figure}
\begin{center}
\includegraphics[angle=0, width=0.5 \textwidth, height=0.7\linewidth] {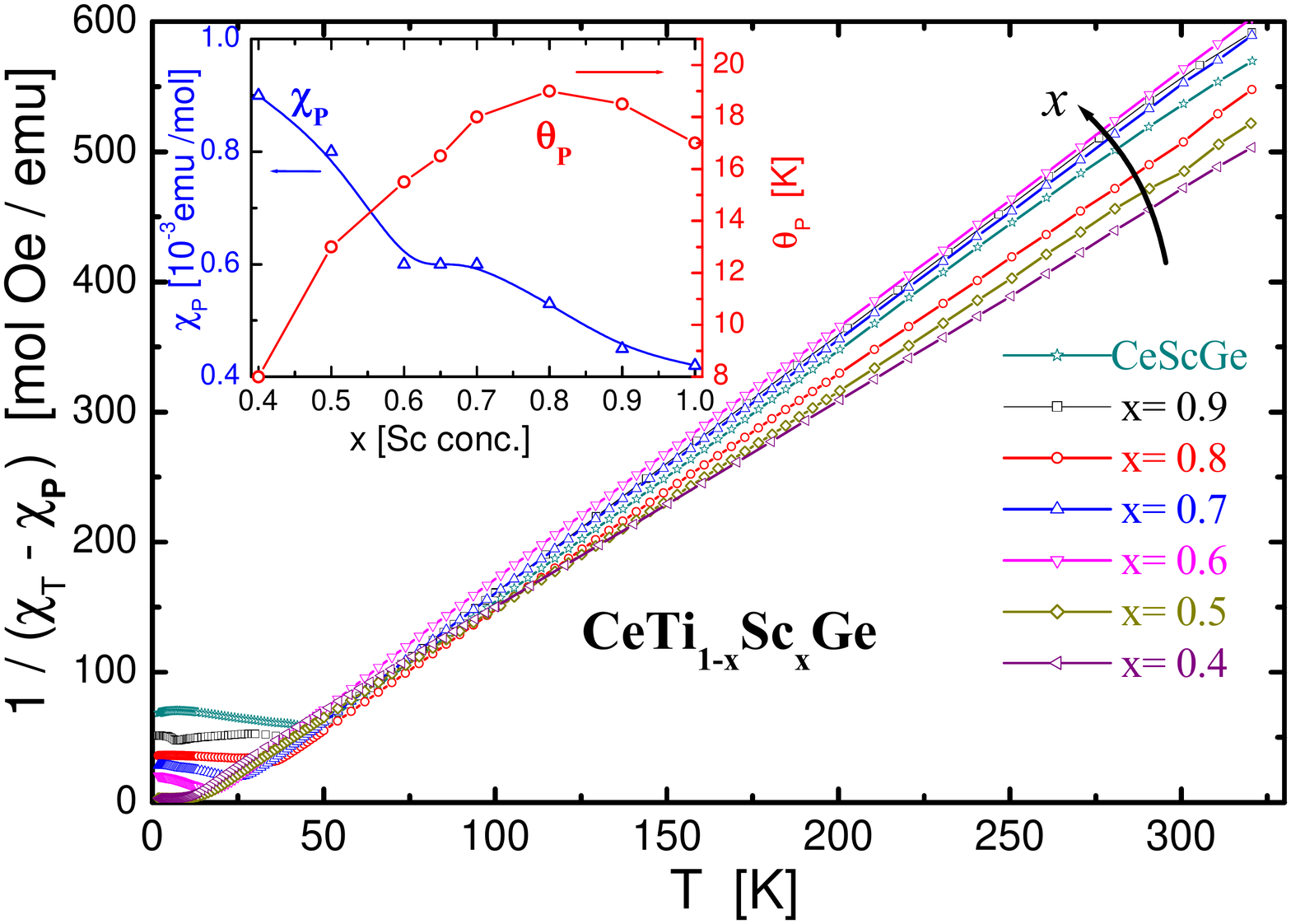}
\end{center}
\caption{(Color on line) Inverse high temperature magnetic
susceptibility measured in a field of $H=10 kOe$, after
subtracting a Pauli type ($\chi_P$) contribution. Inset:
$\chi_P(x)$ contribution and paramagnetic temperature
$\theta_P(x)$. } \label{F3}
\end{figure}

In Fig.~\ref{F3}, the inverse of the high temperature magnetic
susceptibility is presented after subtracting a Pauli type
$\chi_P$ paramagnetic contribution. At high temperature, $\chi
(T)$ is characterized by a decrease of the effective magnetic
moment $\mu_{eff}(x)$ from $\approx 2.25\mu_B$ at $x = 0.4$ to
$\approx 2\mu_B$ at $x = 0.5$ and remains nearly unchanged for
higher Sc content.

The paramagnetic temperature $\theta_P (x)$ is always positive and
increases with Sc concentration from $\theta_P \approx 8K$ at $x =
0.4$ up to 19K at $x = 0.8$, where it starts a slight decrease as
seen in the inset of Fig.~\ref{F3}. A significant Pauli-type
$\chi_P (x)$ contribution is observed along the concentration
range, decreasing from $\chi_P = 0.9 10^{-3}$ at $x = 0.4$ down to
$0.47 10^{-3}$emu/Oe\,mol at $x = 1$. This type of contribution
was already reported for CeScGe \cite{Uwatoko}.

\begin{figure}
\begin{center}
\includegraphics[angle=0, width=0.5 \textwidth, height=0.7\linewidth] {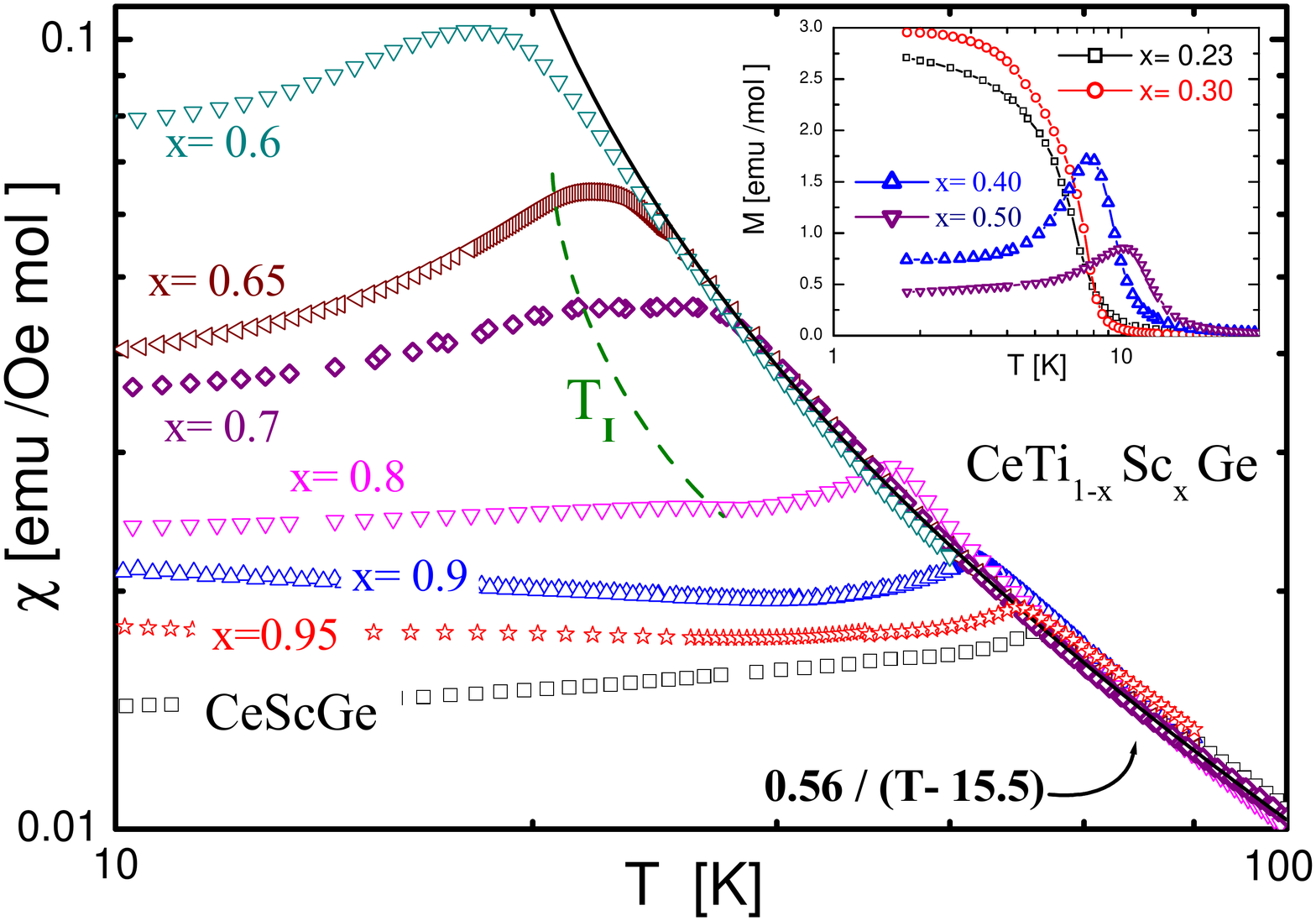}
\end{center}
\caption{(Color online) Low field ($H = 100 Oe$) magnetic
susceptibility in a double logarithmic representation. Full curve
is a reference for the paramagnetic phase, dashed curve indicates
the position of the second transition at $T = T_I$. Inset $M(T)$
measurements for the $x = 0.4$ samples.} \label{F5}
\end{figure}

The magnetic susceptibility results in the range ($T<70$K) are
shown in Fig.~\ref{F5} in a double logarithmic representation. In
the inset, the samples with $x\leq 0.5$ are included using a more
extended $M(T)$ scale. Except for the alloys with $x\leq 0.30$,
all samples show a maximum in $\chi(T)$ at the ordering
temperature. The thermal dependence on the paramagnetic phase is
compared in the figure with a general Curie-Weiss function
$\chi(T) = 0.56 /(T-15.5)$emu/Oe\,mol, including a positive
$\theta_P=15.5$K value, indicating a similar behavior for all
concentrations above the transition.

On the contrary, the ordered phase shows a quasi continuous
variation of the magnetic characteristics with Sc concentration
increase. The typical susceptibility dependence of a FM is
observed in the $x=0.25$ and 0.3 alloys, whereas those samples
with $x\geq 0.4$ show an incipient AFM component, see the inset in
Fig.~\ref{F5}. Except for the former two FM alloys (i.e. $x=0.25$
and 0.3), the magnetic structure seems to be complex and therefore
we will label those magnetic transitions as $T_{ord}$, discussing
the respective magnetic characteristics in the different
concentration regions.

Those regions can be sorted into three ranges: i) for $x\leq 0.6$
the $M(T)$ dependence indicate an increasing mixture of FM and AFM
components in the GS. In this region, the ordering temperature
ranges between $T_{ord}= 7$K for $x = 0.25$ and 19K for $x =
0.65$. At $x=0.65$ the $\chi(T)$ maximum displays two very close
shoulders which can be distinguished after a detailed analysis of
the $\chi(T)$ curvature, i.e. its second derivative
$\partial^2\chi / \partial T^2$. This feature reveals the $x =
0.65$ concentration located very close to a critical point, above
which two branches of the phase boundary are observed. ii) Between
$0.65 \leq x \leq 0.8$, a slight kink is observed at $T = T_I (x)$
below the ordering temperature (see Fig.~\ref{F5}). In this range
of concentration, $T_{ord}$ increases from 19K at $x = 0.65$ up to
35K at $x = 0.8$, whereas $T_I(x)$ increases from 18K up to 26K.
iii) Above $x = 0.9$, $T_{ord}$ increases from 38K at $x = 0.9$ up
to 47K at $x = 1$, whereas $T_I(x)$ increases moderately from 35K
up to 45K, but the associated anomaly smears out.

\subsection{Electrical resistivity}

\begin{figure}
\begin{center}
\includegraphics[angle=0, width=0.5 \textwidth, height=0.7\linewidth] {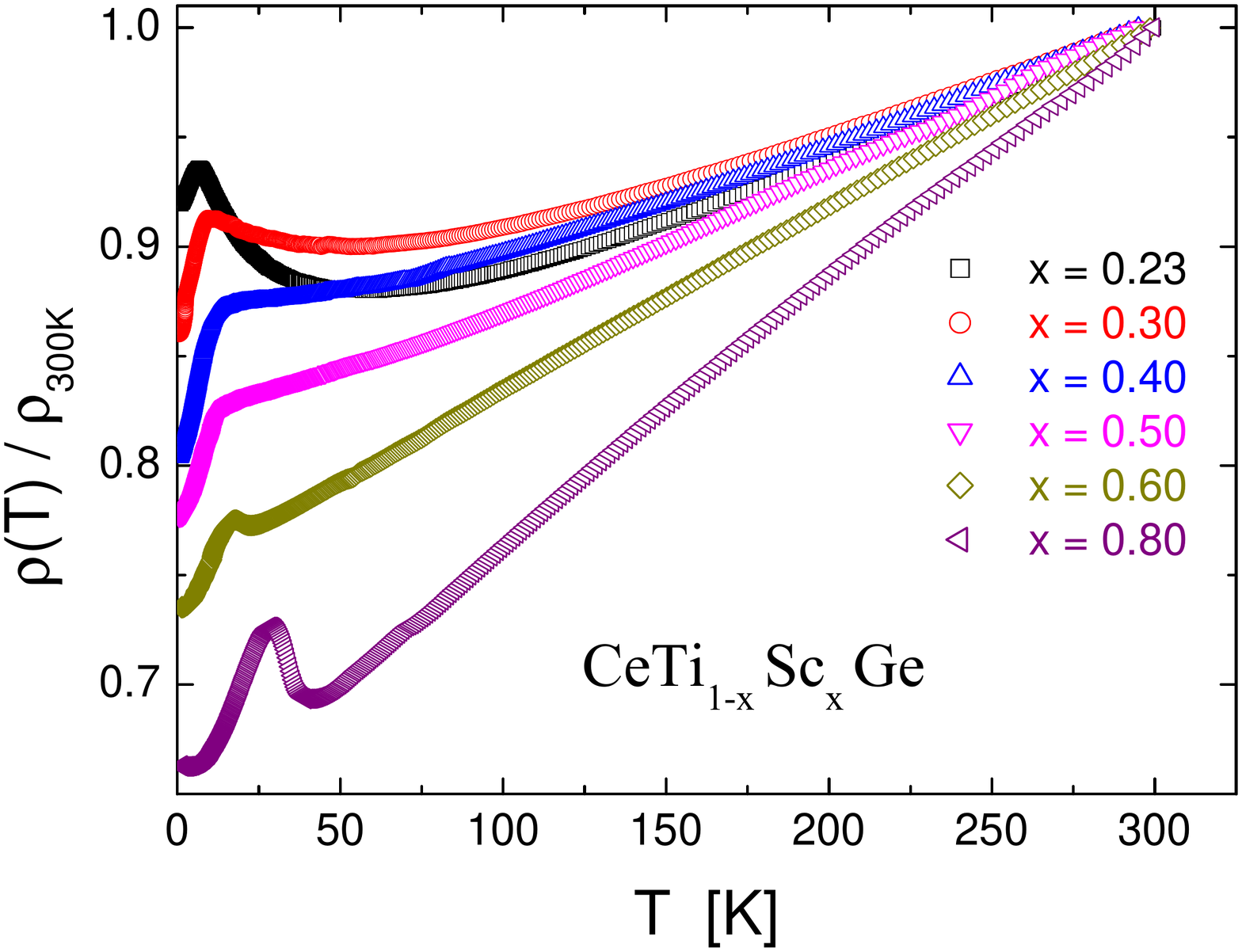}
\end{center}
\caption{(Color online) Electrical resistivity normalized at 300K
of samples between $0.23 \leq x \leq 0.8$.} \label{F4}
\end{figure}

The electrical resistivity measurements normalized at 300K of some
representative concentrations between $0.23 \leq x \leq 0.8$ are
presented in Fig.~\ref{F4}. The increase of $\rho(T)$ approaching
the ordering temperature in $x=0.23$ progressively transforms into
a kink at $x = 0.5$ showing the coherence effect in the ordered
phase. For higher Sc concentrations an anomaly related to an
antiferromagnetic AFM transition the characteristic gap is
observed.

\subsection{Specific heat}

\begin{figure}
\begin{center}
\includegraphics[angle=0, width=0.5 \textwidth, height=0.8\linewidth] {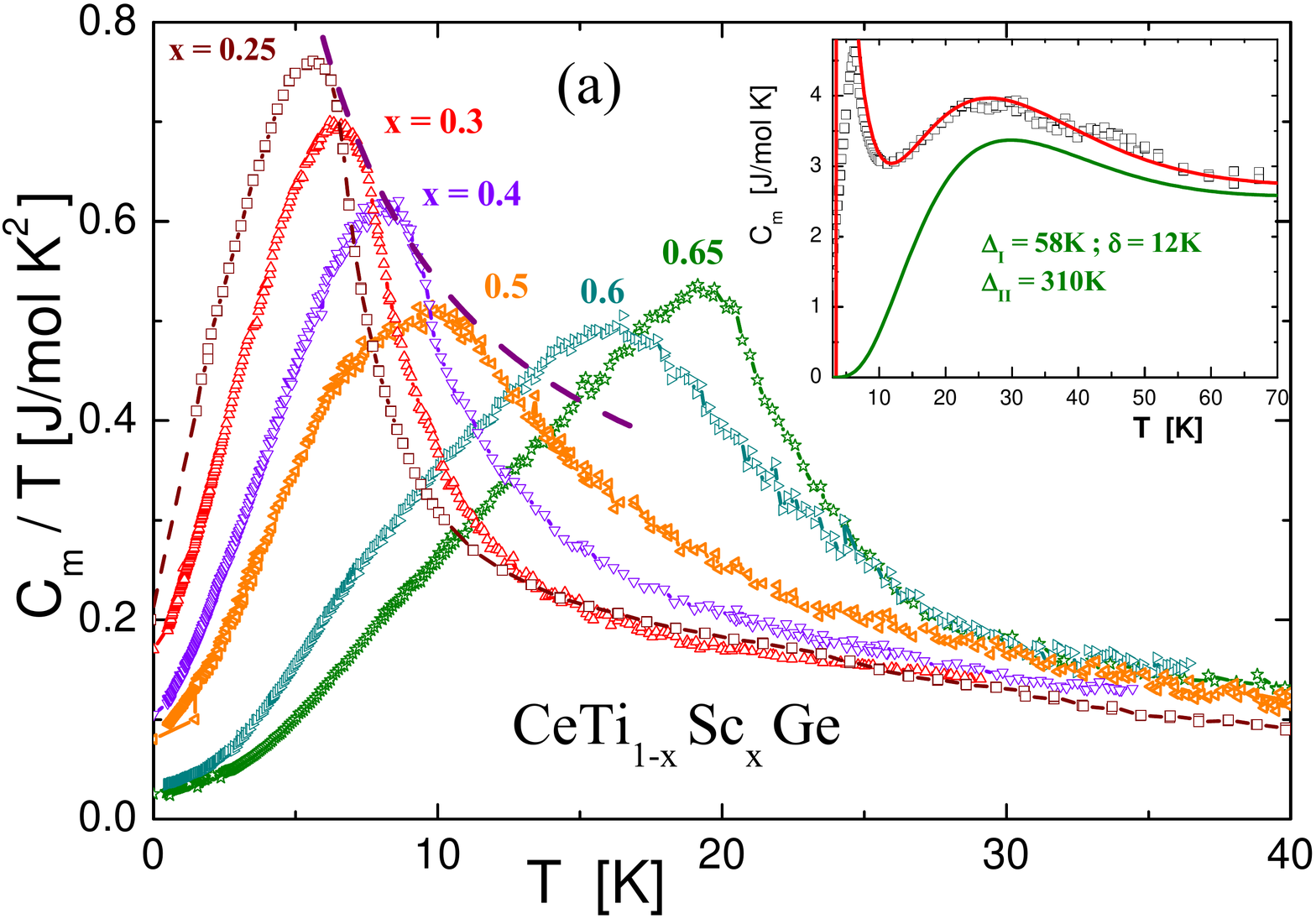}
\end{center}
\caption{(Color online) Magnetic contribution to the specific heat
divided temperature within the 0.25 = x = 0.65 range. Doted curve
indicates the divergent trend of the maximum below x = 0.5. Inset:
Fit of $C_m(T)$ for sample $x=0.25$ at high temperature to extract
the CEF spectrum, see the text.} \label{F7}
\end{figure}

Several peculiar features are observed in the specific heat
results along the concentration variation in coincidence with the
different regimes detected with $M(T)$ measurements. The results
obtained from the measurements performed on the $0.25 \leq x \leq
0.65$ alloys are presented in Fig.~\ref{F7} as $C_m/T$, where
$C_m$ indicates the magnetic contribution to the specific heat
after phonon subtraction. Within the experimental dispersion the
maxima of $C_m(x)/T$ coincide with the respective transition
temperatures extracted from magnetic measurements. A common
feature along this concentration range is the lack of a $C_m$ jump
at $T=T_{ord}$. Instead, a tail in $C_m/T(T>T_{ord})$ is observed,
indicating the presence of magnetic fluctuations above the
transition temperature.

\begin{figure}
\begin{center}
\includegraphics[angle=0, width=0.5 \textwidth, height=1.4\linewidth] {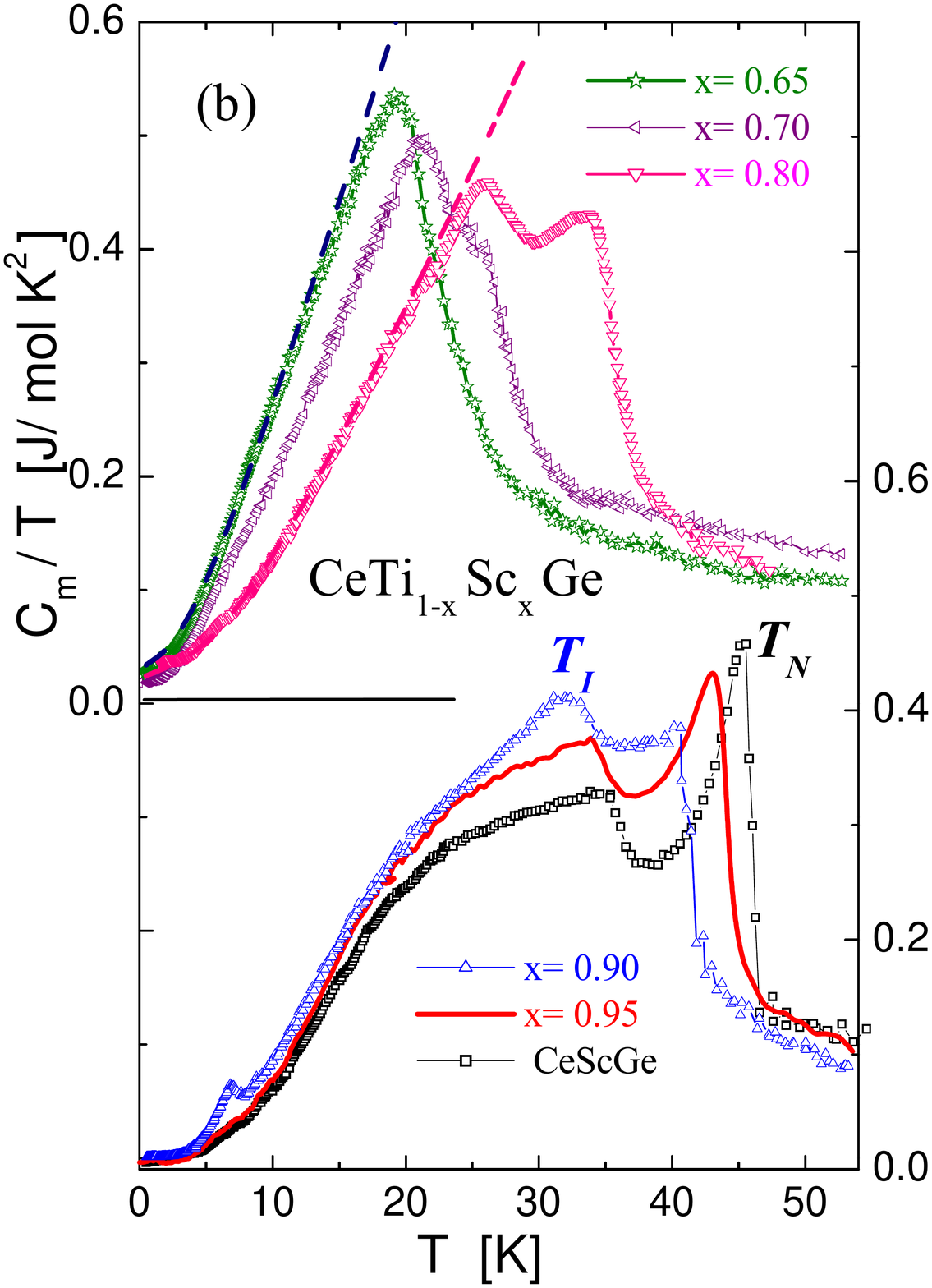}
\end{center}
\caption{(Color online) Magnetic contribution to the specific heat
in the $0.65 \leq x \leq 1$ range, with the $0.90 \leq x \leq 1$
samples shifted for clarity (right axis). Dashed curves represent
the fits on the ordered phase of samples $x=0.65$ and 0.80. }
\label{F8}
\end{figure}

Notably, in the low Sc concentration samples ($0.25\leq x \leq
0.5$) the maximum of $C_m/T$ decreases as $\propto 1/T_{ord}$.
This fact implies that the maximum of $C_m(T_{ord})$ does not
extrapolate to zero with $T_{ord}\to 0$ in contradiction with the
thermodynamic principles. Such a sort of entropy bottleneck would
have impede $T_{ord}\to 0$ even if the CeScSi-type structure would
have extended to $x=0$. Similar behavior was observed in the
Ce$_2$(Ni$_{1-x}$Pd$_x$)$_2$Sn system on the Ni-rich side
\cite{PhilMag} that also shows a structural transition limiting
the $C_m/T$ divergence. Between $x=0.5$ and 0.65, a different
regime sets on because the maximum of $C_m/T$ remains practically
unchanged. Another characteristic of this range of concentration
is the decrease of $C_m/T(T\to 0)=\gamma_0$ extrapolation from
$\gamma_0 \approx 0.22$J mol$^{-1}$K$^{-2}$ to $\approx 0.02$J
mol$^{-1}$K$^{-2}$ between $x=0.25$ and 0.65. Coincidentally, the
increase of the low temperature curvature in $C_m/T(T)$ indicates
a progressive opening of a gap of anisotropy ($\Gamma$) in the
magnon spectrum that for $x=0.65$ reaches $\Gamma=7$K as shown by
the fitting curve included in Fig.~\ref{F8}. For such a fit we
have used the function: $C_m/T= \gamma_0 + B*T*exp(-\Gamma /T)$.
The tail of $C_m/T(T>T_{ord})$ become more extended with
increasing Sc concentration. However, sample $x=0.65$ recovers a
significant slope above $T_{ord}$ indicating a change in the
nature of the transition associated to a reduction of magnetic
fluctuations in the paramagnetic phase.

Between $x=0.65$ and 0.80 the temperature dependence of $C_m/T$ in
the ordered phase can be described with the $C_m/T\propto
T*exp(-\Gamma /T)$ dependence, as shown in Fig.~\ref{F8} for
samples $x=0.65$ and 0.80. For the latter composition the computed
gap is $\Gamma = 15$K.

The main difference with the alloys with lower Sc concentration is
the split of $C_m/T$ into two maxima, in agreement with the
$\chi(T)$ results. While the lower transition (at $T=T_I$) shows a
cusp-like anomaly between $x=0.7$ and 0.9, the upper one
identified as $T_N$ from $\chi(T)$ measurements is associated to a
jump in $C_m/T(T_N)$ within this range of concentration. The cusp
at $T=T_I$ can be related to the vicinity of the critical point
accounting that in these policrystalline samples any eventual
first order transition would be broadened in temperature because
of the random direction distribution of crystals and some
broadness in concentration. Within this range of concentration,
the shape of the upper transition at $T_N$ is typical for a second
order transition.

For $x\geq 0.9$, a further change in $C_m/T(T)$ is observed as
depicted in the lower part of Fig.~\ref{F8}. The broad shoulder of
$C_m/T(T)$ in samples $0.9\leq x \leq 1$ can be attributed to the
increase of the GS degeneracy from $N_{eff} =2$ to 4 as the
contribution of the first exited CEF doublet set on. Also the
transitions change their characteristics because while the one at
$T_I$ transforms into a step like anomaly, that at $T_N$ becomes
sharper and grows significantly. Notice that the $\Delta C_m(T_N)$
jump in $x=1$ is $\approx 16$J mol$^{-1}$K$^{-1}$ is between the
values expected for a doublet ($\Delta C_m =1.5 R$) and a quartet
($\Delta C_m =2.2 R$) predicted in the mean field approximation
\cite{Morphol}. The small anomaly observed in $x=0.9$ at $T\approx
7$K can be attributed to an extrinsic contribution of a small
amount of Ce-oxide.

\section{Discussion}

\subsection{Magnetization}

\begin{figure}
\begin{center}
\includegraphics[angle=0, width=0.5 \textwidth, height=1.4\linewidth] {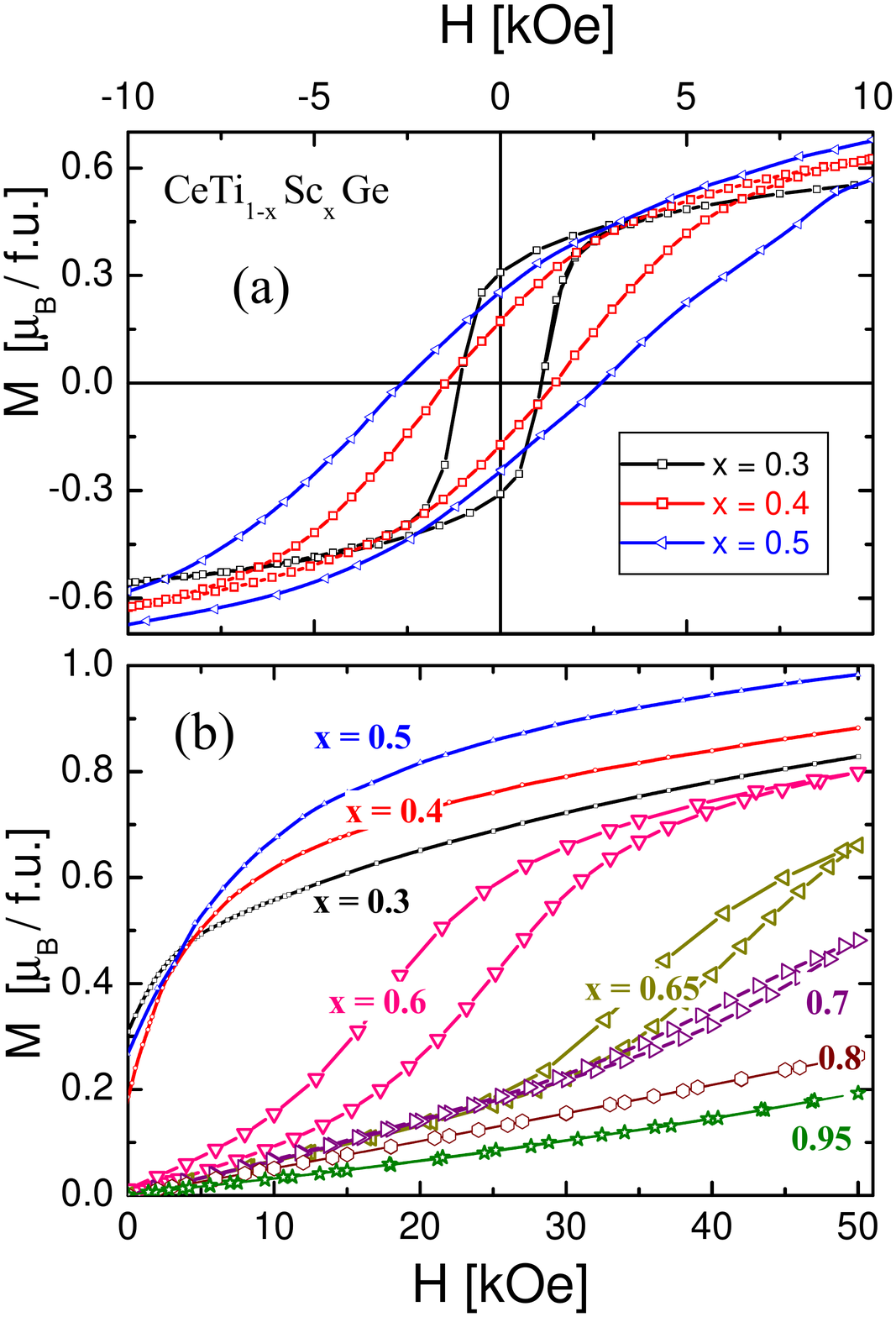}
\end{center}
\caption{(Color online) Magnetization measurements at $T = 1.8$K:
(a) hysteresis loops centered at $H = 0$ for the $0.3 \leq x \leq
0.5$ alloys. (b) Magnetization curves up to $H = 50$kOe, showing a
metamagnetic transformation in samples with $x \geq 0.6$.}
\label{F6}
\end{figure}

The $M(H)$ hysteresis loops measured at $T=1.8$K on the alloys
with $x\leq 0.5$ reveal the FM character of the ordered phase (see
Fig.~\ref{F6}a) concomitant with the increasing value of $M(x)$
measured at $H=50$kOe presented in Fig.~\ref{F6}b. The saturation
magnetization, extracted from Fig.~\ref{F6}b as $M(x,H)=
M_{sat}\times (1-a/H)$, increases from $1.04\mu_B/f.u.$ for
$x=0.3$ up to 1.15$\mu_B/f.u.$ for $x=0.5$ and then decreases to
1$\mu_B/f.u.$ for $x=0.6$. Similarly, the coercive field increases
from $1.1$kOe for $x=0.3$ up to 2.6kOe for $x=0.5$, whereas for
$x\geq 0.6$ a metamagnetic transition occurs. The critical field
$H_{cr}$ increases with concentration up to our experimental limit
of $H=50$kOe with an initial ratio of $\partial H_{cr}/\partial x
= 2.2$kOe/Sc$\%$. However $M(H)$ measurements, performed on
stoichiometric CeScGe \cite{Uwatoko,Singh}, report the
metamagnetic transformation around $H_{cr} \approx 60$kOe with a
weak associated hysteresis. Measurements at higher magnetic field
(up to 30 T \cite{Uwatoko}) reveals anothe magnetic rearrangement
at $H_{crII} \approx 120$kOe. Notably, the area of the hysteresis
loop decreases as $H_{cr}(x)$ tends to its upper value, see
Fig.~\ref{F6}b, and becomes quasi irrelevant at the CeScGe limit
\cite{Singh}.

\subsection{Evaluation of the Crystal Electrical Field splitting}

In order to evaluate the energy of the excited CEF levels, we have
analyzed the $C_m(T)$ dependence of the $x=0.25$ sample which
shows the lowest ordering temperature. For that purpose we have
fit the experimental data accounting that the CEF splits the six
fold Hund's rule multiplet originated in the $J=5/2$ orbital
momentum of Ce into three doublets. Due to the hybridization
(Kondo effect) acting on the first excited level, the standard
Schottky anomaly ($C_{Sch}$) cannot describe the $C_m(T)$
dependence properly. Since a proper fit including usual
hybridization effects ($V_{cf}$) between conduction and $4f$
states acting on each excited level requires complex calculation
protocols \cite{Aligia13}, a simple criterion to mimic the
broadening of the levels was applied in this case. In this
procedure the first CEF exited doublet at $\Delta_I$ is described
using three single Dirac levels equally distributed in energy
around the nominal value of the non hybridized level. This
procedure requires the strength of the hybridization to be smaller
than the CEF splitting ($\Delta_I$), i.e. $V_{cf} < \Delta_I$. For
simplicity, the second excited CEF level (at $\Delta_{II}$) was
included without accounting for hybridization effects because
$\Delta_{II}$ largely exceeds our temperature range of analysis.
The applied formula is:
\begin{equation}
C_{Sch}(T)= \Sigma_i A_i*[(\Delta_i/T)/\cosh(\Delta_i/T)]^2
\end{equation}

The GS contribution was included after fitting the tail of $C_m$
at $T>T_{ord}$ with an arbitrary f(a/T) function, being the total
contribution to the specific heat: $C_{tot} = f(a/T) +
C_{Sch}(T)$. In the inset of Fig.~\ref{F7} the result of this fit
to the experimental data $C_{tot}$ is shown, including the detail
of the $C_{Sch}$ function. The obtained parameters are
$\Delta_I/=35$K and $\Delta_{II}=155$K, with a broadening of the
first CEF level $\delta = 12$K. This procedure was checked with
the evaluation of the magnetic entropy ($S_m$), which for the two
excited doublets reaches $R\ln3$.

This procedure cannot be applied for higher Sc concentrations
because of the increase of $T_{ord}$. Although no significant
changes are expected in the CEF levels splittings, the reduction
of the electrical charges in the transition metal crystal sites
from rich Ti [Ar4s$^2$3d$^2$] to Sc [Ar4s$^2$3d$^1$] may reduce
the strength of the CEF and reduce the $\Delta_{\rm I}$. This
variation may be at the origin of the increasing contribution of
the first excited CEF level to the magnetic ordering.

\subsection{Entropy}

\begin{figure}
\begin{center}
\includegraphics[angle=0, width=0.5 \textwidth, height=0.8\linewidth] {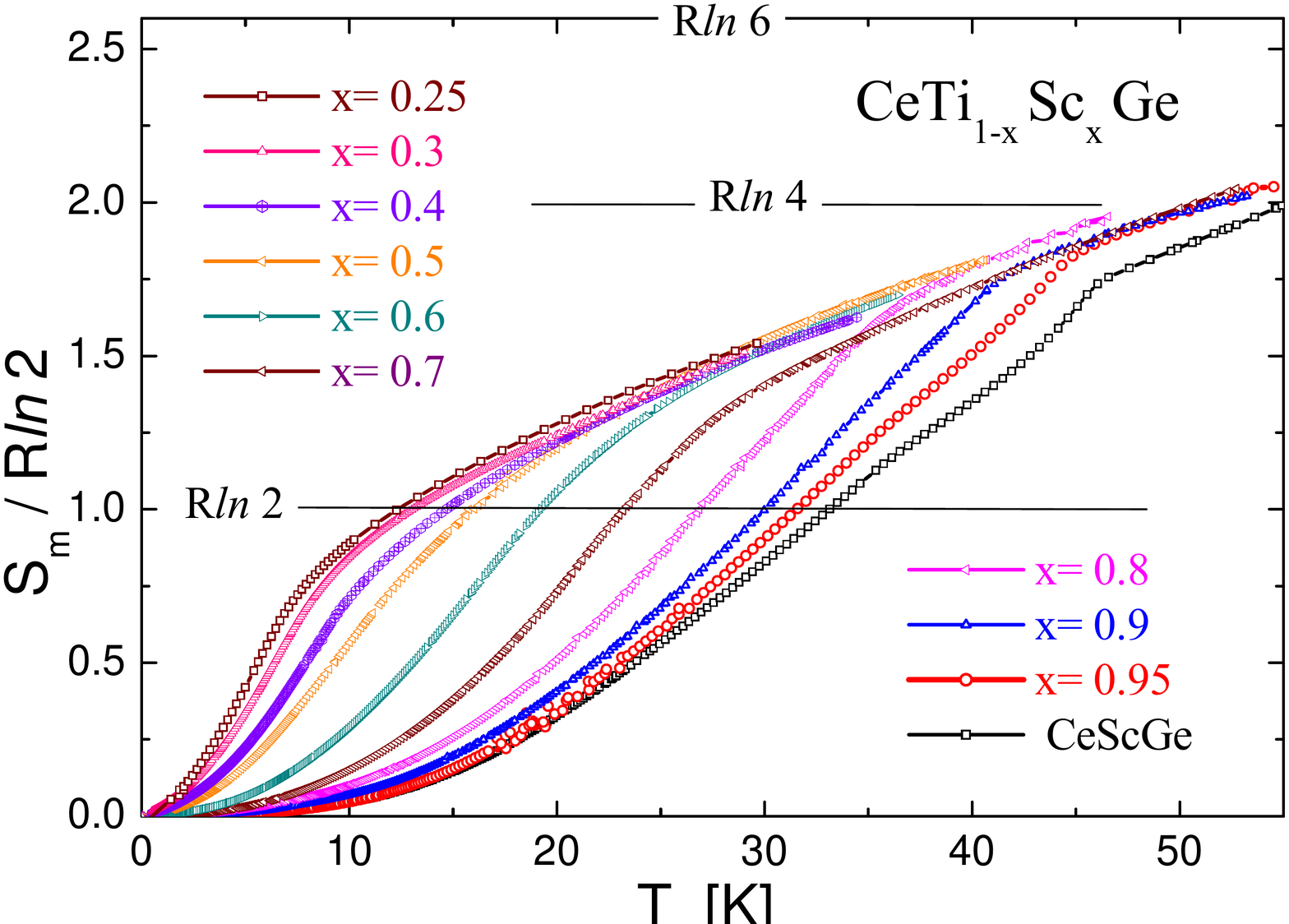}
\end{center}
\caption{(Color online) Thermal evolution of the magnetic
contribution to the entropy normalized by Rln2.} \label{F9}
\end{figure}

The analysis of the thermal evolution of the magnetic contribution
to the entropy $S_m(T)$ is shown in Fig.~\ref{F9} normalized by
the value of a Kramer's doublet level: $R\ln 2$. This parameter
provides relevant information to understand this magnetically
complex system better.

The alloy $x = 0.25$ with the lowest ordering temperature at $T_C
= 7$K reaches $S_m= R\ln 2$ slightly above $T_C$, at $T = 11$K,
suggesting that only the GS doublet contributes to the magnetic
order. However, since in the paramagnetic phase $S_m(T)$ increases
continuously (i.e. without showing any plateau around $R\ln 2$)
one may infer that the tail of the first excited CEF level also
contributes to the entropy at low temperature. This fact confirms
the CEF level spectrum extracted from the fit of $C_m(T)$
performed on this, presented in the inset of Fig.~\ref{F7}.

Although the contribution of that level may be marginal in the
alloys with low Sc content, it becomes significant for higher
concentrations as $S_m (T_{ord})$ increases with $T_{ord}(x)$.
This feature becomes evident for $x = 0.5$ alloys, where $S_m
(T_{ord})$ clearly exceeds $R\ln 2$ and, for $x=1$ practically
reaches the value for two doublets, i.e. $R\ln 4$, in agreement
with the corresponding value observed in $\Delta C_m(T_N)$.

The comparison of the thermal distribution of the entropy above
and below $T_{ord}$ also provides a hint for the effective
dimensionality of the magnetic system. According the theoretic
predictions \cite{ProgPhys}, in two dimensional (2D) systems, the
entropy accumulated between $T=0$ and $T=T_{ord}$ is similar to
that contained in the tail of $C_m(T>T_{ord}$. In this case,
samples with $0.25\leq x \leq 0.5$ Sc concentration fulfil that
property once the $\gamma_0$ contribution is subtracted. Such is
not the case for the samples beyond the critical point for which
the $T_N$ transition shows the characteristic $\Delta C_m(T_N)$
jump of a 3D second order transition.

\subsection{Magnetic phase diagram}

\begin{figure}
\begin{center}
\includegraphics[angle=0, width=0.5 \textwidth, height=0.7\linewidth] {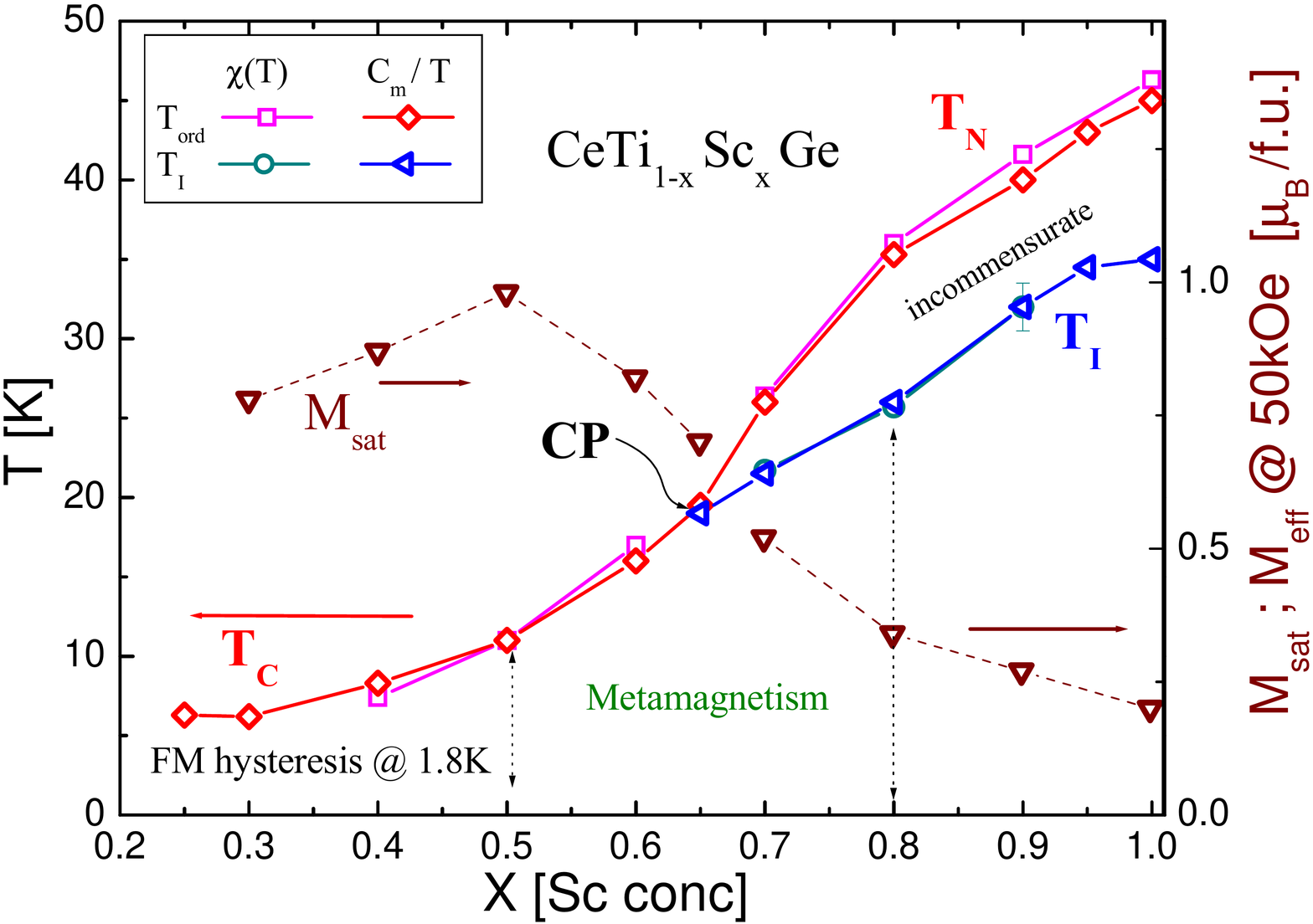}
\end{center}
\caption{(Color online) Sc concentration dependence of the
Magnetic phase diagram showing the transition temperatures
extracted from magnetic and thermal measurements (left axis)
within the range of the CeScSi-type structure. Saturation
magnetization $M_{sat}$ (right axis) of samples below the critical
point CP ($x=0.65$), compared with the effective magnetic moment
$M_{eff}$ of alloys beyond the CP, measured at $T=1.8$K and$\mu_0
H = 5$Tesla.} \label{F10}
\end{figure}

The main parameters of this system are collected in the magnetic
phase diagram presented in Fig.~\ref{F10}. The most relevant
feature is the critical point CP determined at $x \approx 0.65$,
where an intermediate phase sets in between the $T_N(x)$ and
$T_I(x)$ phase boundaries. On both sides of that CP two different
regions were identified. Below $x = 0.65$, a change in the GS
magnetic properties is observed around $x = 0.5$ between a FM GS
(determined by $M vs H$ hysteresis loops of Fig.~\ref{F6}a) and
the region where a metamagnetic transition arises in the field
dependence of the magnetization. In this region ($0.5\leq x \leq
0.7$) a critical field ($H_{cr}$) of the spin flip increases
proportionally to Sc content up to the limit of our applicable
magnetic field of $H = 50$kOe.

The saturation magnetization $M_{sat}$ extrapolated from
Fig.~\ref{F6}b reaches a maximum value at $x=0.5$. These values
are compared in Fig.~\ref{F10} with those obtained from
$M$($H=50$kOe) at $T = 1.8$K of samples with $x\geq 0.7$, that
keep decreasing. Despite of the weakening of the magnetic moments
with Sc concentration, $T_N(x)$ rises continuously up to $x = 1$.
According to the literature \cite{DeLong91}, this behavior is
observed in itinerant systems where SDW develops, in agreement
with the $\rho(T)$ anomaly detected at $T\approx T_{ord}$ in $0.6
\leq x \leq 1$ alloys.

\section{Conclusions}

The main characteristic of this system is the rapid variation of
the magnetic nature of its ordered phase with Sc concentration,
likely due to an increasing role of the first excited CEF doublet
into the GS properties. An encompassing description for this
peculiar system can be proposed as follows:

i) between $x= 0.25$ and 0.5, the local $4f$ moments order FM with
intra-planes interaction involving two consecutive Ce planes that
maximizes the $M_{sat}$ value. Within this range of concentration,
the comparison of the entropy distribution above and below
$T_{ord}$ suggests a 2D magnetic character \cite{ProgPhys}.

ii) Around $x=0.6$, the first excited CEF level becomes more
relevant in the definition of the magnetic GS properties, being
responsible for the change from FM to a AFM character of the
magnetic structure. Nevertheless, such a nearly continuous
variation indicates that both configurations are quite close in
energy as confirmed by the metamagnetic transitions.
Coincidentally, as the AFM character arises, $M_{sat}$ starts to
decrease.

iii) The most significant modification of the GS properties occurs
at the critical point at $x\approx 0.65$. Beyond this
concentration, the phase transition at $T_N$ looks more likely one
of second order type in a 3D context induced by the onset of an AF
interaction inter-planes. An intermediate phase arises below
$T_N(x)$ with evidences for an incommensurate spin density wave
character. The evolution of the specific heat anomaly at the
$T=T_I$ transition suggests a small discontinuity in the magnetic
wave vector. These observation coincide with the fact that the
transition at $T=T_I$ is hardly seen in $\chi(T)$ measurements
because of the small change produced in the AFM structure.

iv) For $x\geq 0.9$ the $C_m/T(T)$ anomaly at $T=T_I$ becomes even
weaker transforming into a second order type whereas the upper one
increases up to rich a $\Delta C_m/T(T_N)$ jump expected for a
four fold degenerate state $\approx 18 J mol^{-1}K^{-1}$. The
itinerant character of this magnetic GS is suggested by the
anomaly in the $\rho(T)$ dependence at $T\approx T_N$ and the
reduced value of the magnetic moment compared with the extremely
high ordering temperature as it occurs in many U compounds
\cite{DeLong91}.

In conclusion, this work provides experimental evidences for this
peculiar system that undergoes a complex transformation in its GS
magnetic properties associated to an increasing contribution of
the first excited CEF level. Beyond a critical point, the
formation of a quasiquartet enhances the exchange interaction
between Ce-planes with the consequent growth of the ordering
temperature. Similar behavior was recently observed in CeCoSi
under pressure showing a high temperature magnetic order also
attributed to a quasiquartet level formation \cite{CeCoSi}. This
facts are ascribed into a scenario of a change in the magnetic
dimensionality from 2D type to 3D and the appearance of spin
density waves with incommensurate wave vector.

\end{document}